\documentclass[preprint,showpacs,preprintnumbers,nofootinbib]{revtex4}
\usepackage{graphicx}% Include figure files
\usepackage{epstopdf}
\usepackage{dcolumn}% Align table columns on decimal point
\usepackage{bm}% bold math

\newcommand{\nn}{\nonumber}

\newcommand{\be}{\begin{eqnarray}}
\newcommand{\ee}{\end{eqnarray}}
\catcode`\@=11
\def\lsim{\mathrel{\mathpalette\@versim<}}
\def\gsim{\mathrel{\mathpalette\@versim>}}
\def\@versim#1#2{\vcenter{\offinterlineskip
\ialign{$\m@th#1\hfil##\hfil$\crcr#2\crcr\sim\crcr } }}
\catcode`\@=12

\def\thefootnote{\fnsymbol{footnote}}

\begin{document}

\def\thefootnote{\fnsymbol{footnote}}
\begin{flushright}
KANAZAWA-11-18\\
October, 2011
\end{flushright}
\vspace*{0.5cm}

\begin{center}

{\Large \bf 
Impact of Inert Higgsino Dark Matter}

\end{center}

\vspace{0.2cm}

\begin{center}
{\large Mayumi Aoki\footnote[1]{e-mail:~mayumi@hep.s.kanazawa-u.ac.jp},  
Jisuke Kubo\footnote[2]{e-mail:~jik@hep.s.kanazawa-u.ac.jp},
Taishi Okawa\footnote[3]{e-mail:~okawa@hep.s.kanazawa-u.ac.jp}  
and
Hiroshi Takano\footnote[4]{e-mail:~takano@hep.s.kanazawa-u.ac.jp}
}
\vspace {0.5cm}\\
{\it Institute for Theoretical Physics, Kanazawa University,\\
        Kanazawa 920-1192, Japan}
\end{center}
\vspace{0.5cm}

{\Large\bf Abstract}

\vspace{0.5cm}
%%%%%%%%%%%%%%%%%Abstract%%%%%%%%%%%%%%%%%%%%%%%%%
We consider  a recently proposed supersymmetric 
 radiative seesaw model which is coupled with the minimal supergravity.
 The conventional R parity and $Z_2$ invariance
 are imposed,  which
ensures the existence of a multi-component dark matter system.
We assume that the pair of
  the lightest  neutralino $\tilde{\chi}$ and the fermionic 
 component $\tilde{\xi}$ of the inert Higgs supermultiplet is dark matter.
If $\tilde{\xi}$ is lighter than $\tilde{\chi}$, 
and the lightest neutral inert Higgs boson is kinematically forbidden to decay
(third dark matter),
the allowed  region
 in the $m_0{\rm \mathchar`-}M_{1/2}$ plane
 increases  considerably,  where $m_0$ and $M_{1/2}$ 
 are the universal soft-supersymmetry-breaking  scalar and gaugino mass, respectively,
 although the dominant component of the multi-component dark matter system  is $\tilde{\chi}$.
There is a wide allowed region  above the recent LHC limit.
 
\newpage
\setcounter{footnote}{0}
\def\thefootnote{\arabic{footnote}}

%\section{Introduction}
The minimal supersymmetric model (MSSM) is one of the attractive extensions
of the standard model (SM) 
\cite{Haber:1984rc}. 
Especially, if it is coupled with  the minimal supergravity
(mSUGRA)\footnote{We use the definition given 
by the Particle Data Group \cite{pdg}.}
the supersymmetry (SUSY) breaking sector is drastically simplified 
\cite{Nilles:1983ge}, and
we have at hand a constrained MSSM (CMSSM) with
the  universal and flavor-diagonal soft-SUSY-breaking parameters.
Because of this simplicity the allowed parameter region in  the SUSY breaking
sector   has become smaller
and smaller as more and more  experimental data 
have become available 
\cite{Buchmueller:2011ki,AbdusSalam:2011fc,Cassel:2011tg,Aad:2011ib,Chatrchyan:2011ek}.
A severe constraint also comes from the relic density of dark matter
\cite{Komatsu:2010fb}
if one assumes that the dark matter candidate is the lightest neutralino
\cite{Griest:1988ma,Jungman:1995df}.
If it is displayed in the $m_0$\,-\,$M_{1/2} $ plane for a given
value of $\tan\beta$, where $\tan\beta$ is the ratio of
 the vacuum expectation value of the 
up-type Higgs field to that of the down-type Higgs field,  $m_0$ and $M_{1/2}$ are
the universal soft scalar  and gaugino mass at the unification scale $M_{\rm GUT}$,
the allowed region is only a narrow strip \cite{Baer:1995nc}.

The constraints on the soft-SUSY-breaking 
parameters  may be relaxed in various ways as 
reviewed  e.g. in \cite{Buchmueller:2011ki,AbdusSalam:2011fc}.
Here we will consider a supersymmetric extension 
\cite{Ma:2006uv,Fukuoka:2009cu} (see also \cite{Ma:2008ba})
of  the model of \cite{Ma:2006km},
in which the neutrino mass 
and mixing are generated  in higher orders of
perturbation theory \cite{Zee:1980ai}.
In a class of recent  radiative seesaw models
the tree-level neutrino mass is protected by an extra  discrete symmetry $Z_2$ 
which, if it is unbroken,  ensures the existence of a $Z_2$ odd stable particle,
a potential candidate for dark matter \cite{Krauss:2002px,Ma:2006km}
\footnote{A variety of similar models have been
recently constructed in 
\cite{Hambye:2006zn,Ma:2007yx,Aoki:2008av,Ma:2008cu,Suematsu:2010gv,Haba:2011cj}.
Leptogenesis 
\cite{Fukugita:1986hr} in radiative seesaw models
has been discussed in \cite{Higashi:2011qq}, and 
baryogenesis \cite{Kuzmin:1985mm} in \cite{Aoki:2008av,Ma:2006fn}.}.
It has been shown  for the model of \cite{Ma:2006km} that 
the lightest right-handed neutrino 
can become a realistic dark matter candidate \cite{Kubo:2006yx} (see also
\cite{Sierra:2008wj}).
A motivating reason to supersymmetrize this type of models is that
because of the extended  Higgs sector 
 the models are meaningful only up to 
energies $\sim$ few TeV \cite{Aoki:2011zg} and
supersymmetry can  considerably improve this situation.
There will  be a set of  potential candidates for dark matter in 
the supersymmetric radiative seesaw models
\cite{Ma:2006uv,Fukuoka:2009cu}, because in addition to $Z_2$, the usual R parity is 
assumed.
To exhaust all such  possibilities is certainly interesting, but it will be beyond
the scope of the present Letter. 
Here we will assume 
that  the lightest neutralino
$ \tilde{\chi}$ and  the lightest inert higgsino $\tilde{\xi}$
are dark matter  candidates.
This combination of the dark matter particles has not been considered 
in the past.
The lightest inert Higgs boson $\xi$, which is 
heavier than $\tilde{\chi}$ and $\tilde{\xi}$ by assumption,
can become the third candidate,
if its decay is kinematically forbidden.
We will show that in the parameter region, where 
the decay of $\xi$ is kinematically not allowed, the  allowed  region
in the $m_0$\,-\,$M_{1/2}$ plane increases  considerably,
if  $\tilde{\xi}$ is lighter than $\tilde{\chi}$.

\vspace{1cm}
%%%%%%
%\section{The model and dark matter candidates}
In the supersymmetric extension \cite{Ma:2006uv,Fukuoka:2009cu} 
of  the model of \cite{Ma:2006km},
a product of  abelian discrete symmetries $R\times Z_2 \times Z_2^L$ 
is assumed to be intact.
The discrete $Z_2$ forbids the tree-level neutrino mass, and $Z_2^L$
is the discrete lepton number, while $R$ is the usual R parity.
The matter content of the model with their quantum numbers
is given in Table I.
${\bf L}$, $ {\bf H}^u, {\bf H}^d$ and $\mbox{\boldmath  $\eta$}^u, \mbox{\boldmath  $\eta$}^d$
stand for $SU(2)_L$ doublets
supermultiplets of the leptons, the MSSM  Higgses and the inert 
Higgses, respectively. The MSSM quarks are ${\bf Q}, {\bf U}^C$ and ${\bf D}^C$ as usual.
Similarly, $SU(2)_L$ singlet
supermultiplets of the charged leptons and right-handed neutrinos are denoted by
${\bf E}^C$ and ${\bf N}^C$.  The gauge singlet supermultiplet 
$\mbox{\boldmath  $\phi$}$ is an additional neutral Higgs supermultiplet 
which is needed
to generate neutrino masses radiatively. 
\begin{table}[t]
\begin{center}
\begin{tabular}{|c|cccccccc|} \hline
 & ${\bf L}$ & ${\bf E}^C$ & ${\bf N}^C $&${\bf H}^u$&${\bf H}^d$& 
 $\mbox{\boldmath  $\eta$}^u $ 
 & $\mbox{\boldmath  $\eta$}^d$ & $\mbox{\boldmath  $\phi$}$ 
  \\ \hline
 $R\times Z_2 $ 
 & $(-,+)$  &  $(-,+)$  &  $(-,-)$
 & $(+,+)$&  $(+,+)$
 &  $(+,-)$ & $(+,-)$& $(+,-)$ 
   \\ \hline
$Z_2^L$ 
 & $-$  &  $-$  &  $-$
 & $+$&  $+$
 &  $+$ & $+$& $+$
   \\ \hline
\end{tabular}
\caption{\footnotesize{The matter content and
the quantum number. $R\times Z_{2}
\times  Z_2^L$ is the unbroken discrete symmetry. 
The quarks of the MSSM are suppressed in the Table.}}
\end{center}
\end{table}
The superpotential is $W=W_Y+W_\mu$, where
\be
W_Y
&=&
Y_{ij}^{u} {\bf Q}_{i} {\bf U}^C_{j}  {\bf H}^u
+Y_{ij}^{d } {\bf Q}_{i} {\bf D}^C_{j} {\bf H}^d
+Y_{ij}^{e} {\bf L}_{i} {\bf E}_{j}^C  {\bf H}^d
+Y_{ij}^{\nu } {\bf L}_{i} {\bf N}_{j}^C \mbox{\boldmath  $\eta$}^u
\nn\\&& +
\lambda^u {\bf H}^d \mbox{\boldmath  $\eta$}^u \mbox{\boldmath  $\phi$}
+\lambda^d \mbox{\boldmath  $\eta$}^d {\bf H}^u \mbox{\boldmath  $\phi$}~,
\label{w1}\\
W_\mu 
&=& 
-\mu_H {\bf H}^u {\bf H}^d
+ \frac{(M_N)_{ij}}{2}{\bf N}_i^C{\bf  N}_j^C
+ \mu_\eta \mbox{\boldmath  $\eta$}^u \mbox{\boldmath  $\eta$}^d
+ \frac{1}{2}\mu_\phi \mbox{\boldmath  $\phi$}^2~.
\label{w2}
\ee
The soft-SUSY-breaking Lagrangian is ${\cal L}_{SB}={\cal L}_{A}+{\cal L}_{m^2}+{\cal L}_{B}$,
where
\be
{\cal L}_{A}
&=&
h_{ij}^{u} {\tilde Q}_{i} {\tilde U}^C_{j}  H^u
+h_{ij}^{d } {\tilde Q}_{i} {\tilde D}^C_{j} H^d
+h_{ij}^{e} {\tilde L}_{i} {\tilde E}_{i}^C  H^d
+h_{ij}^{\nu } {\tilde L}_{i}{\tilde  N}_{j}^C \eta^u
 \nn\\&& 
 +h_{\lambda^u}H^d  \eta^u \phi
+ h_{\lambda^d} \eta^d H^u \phi 
+ h.c.~,
       \label{LSB-A}\\
{\cal L}_{m^2}
&=&
-(\tilde{m}_{Q}^2)_{ij}{\tilde Q}_{i} {\tilde Q}_{j}^*
-(\tilde{m}_{U}^2)_{ij}{\tilde U}_{i}^{C}{\tilde U}_{j}^{C*}
-(\tilde{m}_{D}^2)_{ij}{\tilde D}_{i}^{C}{\tilde D}_{j}^{C*}
-(\tilde{m}_{L}^2)_{ij}{\tilde L}_{i} {\tilde L}_{j}^*
\nn\\& &
-(\tilde{m}_{E}^2)_{ij}{\tilde E}_{i}^{C}{\tilde E}_{j}^{C*}
-(\tilde{m}_{N}^2)_{ij}{\tilde N}_{i}^{C}{\tilde N}_{j}^{C*}
-m_{H^u}^2H^{u} H^{u*}
-m_{H^d}^2H^{d} H^{d*}
\nn\\& &
-m_{\eta^u}^2\eta^{u} \eta^{u*}
-m_{\eta^d}^2\eta^{d} \eta^{d*}
-m_{\phi}^2 \phi \phi^*~,
       \label{LSB-m2}\\
{\cal L}_{B}
&=&
-( B_H H^u H^d + \frac{(B_N)_{ij}}{2}\tilde{{ N}}_i^C\tilde{{N}}_j^C
+ B_\eta \eta^u \eta^d + \frac{1}{2}B_\phi \phi^2 + h.c.)~.
       \label{LSB-B}
\ee
Our notation is such that the component fields with a tilde have  odd R parity.
Since the model is coupled with the mSUGRA,
the  soft-SUSY-breaking parameters are universal  and flavor-diagonal
at $M_{\rm GUT}$ and the underlying  parameters  are
\footnote{Since the matter content of the present model is different
from that of the MSSM,  the unified soft parameters $m_0$ and
$M_{1/2}$ cannot be directly identified with those of the MSSM.
In the renormalization group (RG) running we  add a pair of 
${\bf d}^C$ and $\bar{{\bf d}}^C$ ($\overline{\bf{3}}$ and  $\bf{3}$  of $SU(3)_C$)
to obtain  gauge coupling unification.
So,  for the gaugino mass we have
$M_{1/2}=
M_{1/2}^{\rm MSSM} (\alpha_{\rm GUT}/\alpha_{\rm GUT}^{\rm MSSM})
\simeq 1.2~ M_{1/2}^{\rm MSSM}$.
This change of $M_{1/2}$ can take into account
the dominant change in the RG running of $m_0$ so that
$m_0$ can be approximately identified with $m_0^{\rm MSSM}$.
We assume that ${\bf d}^C$ and $\bar{{\bf d}}^C$  are  $Z_2$ odd, so that
a Yukawa coupling of the form 
${\bf Q}~{\bf d}^C\mbox{\boldmath  $ \eta $}^d $ is possible.
This Yukawa coupling  will contribute to the RG running. Here we assume it is
negligibly  small.  
}
\be
& &m_0~,~M_{1/2}~,~A_0~, \tan\beta~,
~\mbox{sign}(\mu_H)~,~\mu_\eta~,~\mu_\phi~,~M_N~,~B_\eta~,
~B_\phi~,~B_N~.
\label{parameters}
\ee
It is essential for the radiative seesaw model that the discrete symmetry $Z_2$ 
is unbroken. Since our model is coupled with the mSUGRA, it is not obvious
that $Z_2$ is not spontaneously broken at low energy.
We assume that there exists a set of the boundary conditions
at $M_{\rm GUT}$ 
such that $Z_2$ remains unbroken.
Furthermore, the Yukawa couplings  $Y^{\nu}_{ij}$ are
 the important ones for 
the radiative generation of the neutrino mass.
They need not be necessarily small. Therefore, they drive
the soft-SUSY-breaking scalar mass
matrix $\tilde{m}^2_L$ to 
deviate from the universal, flavor-diagonal  form  
so that lepton flavor violations are generated 
at low energy \cite{Barbieri:1995tw}.
Here we assume that we can 
impose  certain constraints on  $Y^{\nu}_{ij}$ to suppress the lepton flavor violations
without having contradictions with the observed neutrino mass and mixing.

There are many candidates for the dark matter in this model 
\cite{Ma:2006uv,Fukuoka:2009cu}.
The lightest combination
of each row in Table II could be a dark matter.
In this Letter we assume that the lightest combination 
(denoted by $\tilde{\xi}$)
of $\tilde{\eta}^{u0}, \tilde{\eta}^{d0},\tilde{\phi}$, 
the lightest combination 
(denoted by $\xi$)
of $\eta^{u0}, \eta^{d0},\phi$ and
the lightest combination  (denoted by $\tilde{\chi}$)
 of $\tilde{h}^{u0}, \tilde{h}^{d0}, \tilde{Z}, \tilde{\gamma}$
are dark matter particles.
So, $\tilde{\chi}$  is  the lightest supersymmetric particle (LSP)  in  the MSSM sector.
Since the relic density of $\tilde{\xi}$,
$\Omega_{\tilde{\xi}} h^2$ for instance, can vary in principle from $0$ to the 
experimentally observed 
value $\simeq 0.11$ \cite{Komatsu:2010fb}, the relic density $\Omega_{\tilde{\chi}} h^2$ 
can also vary in the entire interval  below the maximal value.
Since the spectrum in the inert Higgs sector plays a crucial role in 
evaluating the relic densities, we first discuss
it in some detail before we come to the calculation of 
the relic density of the dark matter particles.
\begin{table}[t]
\begin{center}
{\renewcommand\arraystretch{1.2}
\begin{tabular}{|c|c|c|} \hline
 $R\times Z_2\times Z_2^L $ & Bosons &Fermions \\ \hline
  $(-,+,+)$  &  
  & $ \tilde{h}^{u0},\tilde{h}^{d0}, {\tilde Z}, {\tilde \gamma}$\\ \hline
  $(-,-,+)$ &     &  $\tilde{\eta}^{u0},
  \tilde{\eta}^{d0}, \tilde{\phi}$
           \\ \hline
  $(+,-,+)$ &  $\eta^{u0},\eta^{d0},
 \phi$ &   \\ \hline
  $(-,-,-)$ &  $\tilde{N}^C$ &  \\ \hline
  $(+,-,-)$ &  & $N^C $
     \\ \hline
    $(-,+,-)$  & ${\tilde\nu}_L$ &
       \\ \hline
\end{tabular}
}
\caption{\footnotesize{The dark matter candidates. 
$\eta^{u0}$ and $\eta^{d0}$ are the neutral
scalar components of $\mbox{\boldmath  $\eta$}^u$  and 
$\mbox{\boldmath  $\eta$}^d$, respectively.
The $(+,+,-)$ candidates 
are dropped, because
they are the left-handed neutrinos.}}
\end{center}
\end{table}

\vspace{0.5cm}
 The inert $SU(2)_L$ doublet Higgs supermultiplets $\{\mbox{scalar},\mbox{fermion}\}$
 are defined as
 \be
\mbox{\boldmath  $\eta$}^u &=& \left(\begin{array}{c}
 \{\eta^{u+}~,~ \tilde{\eta}^{u+}\}\\
\{ \eta^{u0}~,~ \tilde{\eta}^{u0} \}
 \end{array}\right)~,~
\mbox{\boldmath  $\eta$}^d = \left(\begin{array}{c}
 \{\eta^{d0}~,~\tilde{\eta}^{d0}\}\\
\{ \eta^{d-}~,~\tilde{\eta}^{d-}  \}
 \end{array}\right)~. \label{doublet}
  \ee
  Further we decompose the neutral fields  into the real and imaginary parts as
$\eta^{u0}  =(\eta^{u0}_R+i \eta^{u0}_I)/\sqrt{2}$,
$\eta^{d0}  =(\eta^{d0}_R+i \eta^{d0}_I)/\sqrt{2}$ and
$\phi  =(\phi_R+i \phi_I)/\sqrt{2}$.
Then the mass matrix of the CP even neutral inert Higgs fields can be written as
 \footnotesize
\begin{eqnarray}
{\bf m}^2_{\xi^0_R} &=&\left(
\begin{array}{ccc}
X_{uu}&  \epsilon B_{\eta} +\lambda^u\lambda^d s_\beta c_\beta v^2/2  &
X_{u\phi}\\
\epsilon B_{\eta} +\lambda^u\lambda^d s_\beta c_\beta v^2/2
&X_{dd} &X_{d\phi}\\
X_{u\phi} &X_{d\phi}&\mu_{\phi}^2 +m^2_{\phi}+\epsilon B_\phi+[(\lambda^u s_\beta)^2+
(\lambda^d c_\beta)^2]v^2/2
\end{array}\right)~
\label{m2-eta0}
\end{eqnarray}
\normalsize
with $\epsilon =1$ in the
$(\eta^{u0}_R, \eta^{d0}_R, \phi_R)$  basis, where
\be
X_{uu} &= & m^2_{\eta^u}+\mu_{\eta}^2- c_{2\beta}M_Z^2/2
+(\lambda^u c_\beta v)^2/2~,\\
X_{dd} &= &m^2_{\eta^d}+\mu_{\eta}^2+c_{2\beta}M_Z^2/2
+(\lambda^d s_\beta v)^2/2~,\\
X_{u\phi} &= &-[(\lambda^d\mu_\eta+\epsilon\lambda^u\mu_H)s_\beta 
+(\lambda^u \mu_\phi-\epsilon h_{\lambda^u})c_\beta]v/\sqrt{2}~,\\
X_{d\phi} &=&
-[(\lambda^u\mu_\eta+\epsilon \lambda^d\mu_H)c_\beta 
+(\lambda^d \mu_\phi-\epsilon h_{\lambda^d})s_\beta]v/\sqrt{2}~,
\ee
and $s_\theta (c_\theta)$ represents $\sin\theta (\cos\theta)$. 
In deriving the mass matrix (\ref{m2-eta0}) we have assumed that all the parameters 
appearing  in (\ref{w1})$\sim$ (\ref{LSB-B}) are real.
The eigenvalues of ${\bf m}^2_{\xi^0_R}$  in ascending order are denoted by 
$m_{\xi_{Ri}^0}^2~
(i=1,2,3) $, and the eigenstates are denoted by $\xi_{Ri}^0$:
\be
\left( \begin{array}{c}
\eta^{u0}_R\\ \eta^{d0}_R \\ \phi_R
\end{array}\right)&=&\left( \begin{array}{ccc}
 & & \\
 & C_{ij}^R&  \\
 &  &  
\end{array}\right)
\left( \begin{array}{c}
\xi^0_{R1}\\ \xi^0_{R2} \\ \xi^0_{R3}
\end{array}\right)~.
\label{c-xi0}
\ee
The mass matrix  ${\bf m}^2_{\xi^0_{I}}$ for the CP odd components 
can be obtained from 
(\ref{m2-eta0})  with $\epsilon=-1$, where we denote their eigenstates by $\xi^0_{Ii}$.
The mass matrix for the charged inert Higgs fields is given by
\begin{eqnarray}
{\bf m}^2_{\xi^+} &=&\left( \begin{array}{cc}
m^2_{\eta^u}+\mu_{\eta}^2- c_{2\beta}(1-2c_W^2)M_Z^2/2 &  -B_{\eta} \\
 -B_{\eta}& m^2_{\eta^d}+\mu_{\eta}^2+c_{2\beta}(1-2c_W^2)M_Z^2/2
\end{array}\right)
\label{m2-eta+}
\end{eqnarray}
in the basis of $(\eta^{u+},~(\eta^{d-})^*)$.
The eigenvalues of (\ref{m2-eta+}) are 
$m_{\xi_1^+}^2$ and $m_{\xi_2^+}^2$
with $m_{\xi_1^+}^2< m_{\xi_2^+}^2$, and the eigenstates are
\be
\left( \begin{array}{c}
\eta^{u+}\\ (\eta^{d-} )^*
\end{array}\right)&=&\left( \begin{array}{ccc}
 & & \\
 & C_{kl}^+&  \\
 &  &  
\end{array}\right)
\left( \begin{array}{c}
\xi^+_{1}\\ \xi^+_{2} 
\end{array}\right)~.
\label{c-xiP}
\ee

Because of the boundary conditions 
$ m^2_{\eta^u}= m^2_{\eta^d}=m^2_{\phi}=m_0^2$
at the unification scale $M_{\rm GUT}$, the soft scalar masses are constrained. 
 At low energy we expect that
$m^2_{\phi} < m^2_{\eta^u} <  m^2_{\eta^d} $,
because $\phi$ is gauge singlet, and
\be
\frac{d(m^2_{\eta^d} -  m^2_{\eta^u} )(\mu)}{d\mu}
&\sim & -\frac{6}{16 \pi^2}m_0^2 ~\mbox{Tr} [Y^\nu (Y^\nu)^T] < 0~.
\ee
Further, neglecting the D-term contributions along with the assumption of
small $\lambda^{u,d}$ and $h_{\lambda^{u,d}}$
 in (\ref{m2-eta0}), we find that the upper bound
of the smallest eigenvalues  of ${\bf m}^2_{\xi^0_{R,I}}$ and  
${\bf m}^2_{\xi^+}$
($m_{\xi^0_{R,I1}}^2$ and
$m_{\xi^+_1}^2$) can be written as
\be
m_{\xi^0_{R,I1}}^2~,~ m_{\xi^+_1}^2&\lsim & 
\frac{1}{2}(m^2_{\eta^u}+m^2_{\eta^d})+\mu_\eta^2-|B_\eta|~.
\ee
Since the soft mass $B_\eta$ is a free parameter, we regard 
$m_{\xi^0_{R1}}^2 (\simeq m_{\xi^0_{I1}}^2 \simeq m_{\xi^+_1}^2)$ as a free parameter.
Under the assumption mentioned above, we may approximately write
the lightest mass eigenstates as
\be
\xi^0_{R1} &\simeq &\frac{1}{\sqrt{2}}(\eta^{u0}_R-\eta^{d0}_R)~,~
\xi^+_1 \simeq \frac{1}{\sqrt{2}}(\eta^{u+}+(\eta^{d-})^*)~,
\label{m2-min}
\ee
and similarly for $\xi^0_{I1}$.
In Table III we summarize the approximate mass eigenstates 
with their approximate mass eigenvalues
squared .
\begin{table}[t]
\begin{center}
{\renewcommand\arraystretch{1.2}
\begin{tabular}{|c|c|c|} \hline
Mass eigenstate & Composition & Mass 
  \\ \hline
 $\xi^0_{R1}$ 
 & $(\eta_{R}^{u0}-~\eta_{R}^{d0})/\sqrt{2}$ &  $m_{\xi^0_{R1}}^2\simeq m_0^2+
 \mu_\eta^2+0.4 M_{1/2}^2-B_\eta$  
   \\ \hline
$\xi^0_{R2}$ 
 &$\phi_{R}$ &  $m_{\xi^0_{R2}}^2\simeq m_0^2+\mu_\phi^2+B_\phi$  
       \\ \hline
 $\xi^0_{R 3}$ 
 & $(\eta_{R}^{u0}+\eta_{R}^{d0})/\sqrt{2}$ &  
 $m_{\xi^0_{R3}}^2\simeq m_0^2+\mu_\eta^2+0.4 M_{1/2}^2+B_\eta$  
   \\ \hline
 $\xi^0_{I1}$ 
 & $(\eta_{I}^{u0}+\eta_{I}^{d0})/\sqrt{2}$ &  $m_{\xi^0_{I1}}^2\simeq m_0^2+
 \mu_\eta^2+0.4 M_{1/2}^2-B_\eta$  
   \\ \hline
$\xi^0_{I 2}$ 
 &$\phi_{I}$ &  $m_{\xi^0_{I2}}^2\simeq m_0^2+\mu_\phi^2-B_\phi$  
       \\ \hline
 $\xi^0_{I 3}$ 
 & $(\eta_{I}^{u0}-\eta_{I}^{d0})/\sqrt{2}$ &  
 $m_{\xi^0_{I3}}^2\simeq m_0^2+\mu_\eta^2+0.4 M_{1/2}^2+B_\eta$  
   \\ \hline
    $\xi^+_{1}$ 
 & $(\eta^{u+}+(\eta^{d-})^*)/\sqrt{2}$ &  $m_{\xi^+_{1}}^2\simeq  m_0^2
 +\mu_\eta^2+0.4 M_{1/2}^2-B_\eta$  
   \\ \hline
       $\xi^+_{2}$ 
 & $(\eta^{u+}-(\eta^{d-})^*)/\sqrt{2}$ &  $m_{\xi^+_{2}}^2\simeq  m_0^2
 +\mu_\eta^2+0.4 M_{1/2}^2+B_\eta$  
   \\ \hline
\end{tabular}
}
\caption{\footnotesize{The mass eigenstates $\xi^0_{R,Ii}$ and $\xi^+_k$ with positive $B_\eta$ and
$B_\phi$. The order changes  depending on the size and sign of
 $B_\eta$ and $B_\phi$.  
The decomposition into the flavor eigenstates and the mass eigenvalues squared are approximate: We neglected the D-term contributions and assumed that the Yukawa couplings
in the lepton sector including $\lambda^u, \lambda^d, h_{\lambda^u}$ 
and $h_{\lambda^d} $
 are small. Here we have assumed that $\xi^0_{R,I1}$ consists mostly from
 $\eta ^{u0}_{R,I}$ and   $\eta ^{d0}_{R,I}$.
$m_0$ and $M_{1/2}$ are the universal soft scalar mass
and the universal gaugino mass at the unification scale, respectively.
(See the comment of footnote 3.)}}
\end{center}
\end{table}

We next come to the inert higgsino sector.
The charged inert  higgsinos $ \tilde{\eta}^{u+}$ and 
$ \overline{\tilde{\eta}^{d-}}$
form a Dirac spinor $\tilde{\xi}^+$ with the mass $M_{\tilde{\xi}^+}=\mu_\eta$.
The mass matrix of the neutral inert higgsinos is given by
\be
{\bf M}_{\tilde{\xi}^0}&=&\left( \begin{array}{ccc}
 0 & \mu_\eta & -\lambda^u c_\beta v/\sqrt{2} \\ 
\mu_\eta & 0&  -\lambda^d  s_\beta v/\sqrt{2}  \\
- \lambda^u  c_\beta v/\sqrt{2}  &  -\lambda^d  s_\beta v/\sqrt{2} & 2\mu_\phi \\
\end{array}\right)~
\label{m1-eta0}
\ee
in the $  (\tilde{\eta}^{u0}, \tilde{\eta}^{d0}, \tilde{\phi}) $ basis.
The mass eigenstates $\tilde{\xi}^0_i$ 
with the mass $M_{\tilde{\xi}_i^0}~(M_{\tilde{\xi}}\equiv M_{\tilde{\xi}_1^0})$  are defined as
\be
\left( \begin{array}{c}
\tilde{\eta}^{u0}\\ \tilde{\eta}^{d0} \\ \tilde{\phi}
\end{array}\right)&=&\left( \begin{array}{ccc}
 & & \\
 & \tilde{C}_{ij}^0&  \\
 &  &  
\end{array}\right)
\left( \begin{array}{c}
\tilde{\xi}^0_{1}\\ \tilde{\xi}^0_{2} \\ \tilde{\xi}^0_{3}
\end{array}\right)~.
\label{c-tilde-xi0}
\ee
The lightest one $\tilde{\xi}=\tilde{\xi}^0_1$ 
is the dark matter candidate and is a linear combination of the form
\be
\tilde{\xi} &=& \tilde{\xi}^0_1=(\tilde{C}_{11}^{0})^* \tilde{\eta}^{u0}
+(\tilde{C}_{21}^{0})^* \tilde{\eta}^{d0}+(\tilde{C}_{31}^{0})^* \tilde{\phi}~.
\label{tilde-xi}
\ee
One can show from (\ref{m1-eta0})
that the  charged inert  higgsino
is always heavier than the dark matter 
$\tilde{\xi}$.

\begin{figure}[htbp]
\begin{center}
\includegraphics*[width=0.6\textwidth]{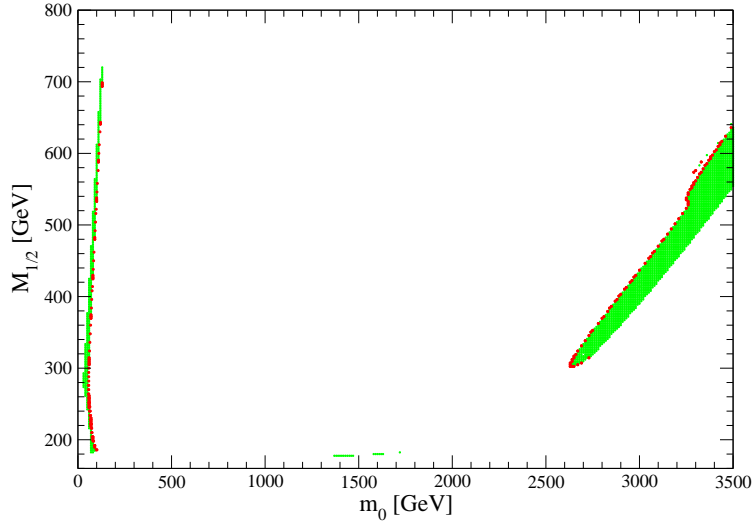}
\caption{\label{cmssm}{\scriptsize
The allowed region in the $m_0$\,-\,$M_{1/2}$ plane.
For the green area the relic density $\Omega_{\tilde{\chi}}h^2$ is assumed to
vary from zero to $0.11$.
The red area corresponds to $\Omega_{\tilde{\chi}}h^2=0.1126\pm 0.0036$ \cite{Komatsu:2010fb}.
We have computed $\Omega_{\tilde{\chi}} h^2$ using 
the package ``micrOMEGAs'' \cite{Belanger:2010gh}
with a set of the input parameters:
 $~A_0=0,~\tan\beta=10$ and $\mbox{sign}(\mu_H)=+$.
Included are the constraints coming from the stau LSP, the electroweak symmetry breaking
and the LEP chargino mass limit.
(See the comment of footnote 3.)}}
\end{center}
\end{figure}

\vspace{0.5cm}
The annihilation rate of  $\tilde{\xi}$ 
depends  on the mixing parameters $\tilde{C}^0_{i1}$ in (\ref{tilde-xi}),
because the dominant contributions to the rate are due to gauge interactions and 
$\tilde{\phi}$ has no gauge coupling.
If $\tilde{\xi}$  is $\tilde{\eta}$-like, the $\tilde{\xi}$ 
behaves like a higgsino-like $\tilde{\chi}$ of the MSSM  so that 
the annihilation cross section tends to be  large. 
It decreases as $\tilde{\xi}$ contains more $\tilde{\phi}$.
So, the annihilation cross section  can be controlled
by $\tilde{C}^0_{i1}$. Therefore,  the relic density
$\Omega_{\tilde{\xi}}h^2$ can vary from a small to the observed value
$\simeq 0.11$ \cite{Komatsu:2010fb}.
That is, $\Omega_{\tilde{\chi}}h^2$, too, may assume a value $\lsim 0.11$.
In Fig.~\ref{cmssm} we plot the allowed region (green) in the
$m_0$\,-\,$M_{1/2}$ plane. The red area is the  region 
for $\Omega_{\tilde{\chi}}h^2 = 0.1126\pm 0.0036$ \cite{Komatsu:2010fb}.
As we see from Fig.~\ref{cmssm} the allowed region in the 
$m_0$\,-\,$M_{1/2}$ plane expands only slightly,
 if the annihilation cross section of $\tilde{\chi}$
into inert Higgs bosons and inert higgsinos  is sufficiently suppressed.

Next we discuss the case that  the fermions  in the 
inert Higgs sector are lighter than $\tilde{\chi}$,
and make therefore the following assumption on the mass hierarchy:
\be
M_{\tilde{\xi}_{i}^0},~M_{\tilde{\xi}^+}&<  & M_{\tilde{\chi}} 
< m_{\xi^0_{R,Ij}},~m_{\xi^+_{k}}~.
\label{hierarchy}
\ee
The annihilation cross section $\sigma_{\tilde{\xi}}$ of $\tilde{\xi} (=\tilde{\xi}_1^0)$  can be obtained
from that of the higgsino-like $\tilde{\chi}$ of the MSSM.
The diagrams are shown in Fig.~\ref{annihilation0}.
As in the case of the MSSM $\sigma_{\tilde{\xi}}$ 
tends to be   large if $\tilde{\xi}$ is $\tilde{\eta}$-like,
where the diagrams (a), (b), (c) and (d) in Fig.~\ref{annihilation0}
are dominant, if the scalar
partner $\tilde{N}^C$ of the right-handed neutrino
is much heavier than $Z$.
Under this assumption  the only possibility to 
suppress $\sigma_{\tilde{\xi}}$ 
is to increase the $\tilde{\phi}$ content in $\tilde{\xi}$
as we have discussed above. 
Note that the relic density $\Omega_{\tilde{\xi}} h^2$ is inversely  proportional
to $( |\tilde{C}^0_{11}|^2+|\tilde{C}^0_{21}|^2)^2
=(1-|\tilde{C}^0_{31}|)^2$
 (if we neglect the contribution (e) in Fig.~\ref{annihilation0}),
 where $\tilde{C}^0_{ij}$ are defined in (\ref{c-tilde-xi0}). 
We have computed $\Omega_{\tilde{\xi}} h^2$  for $\tilde{C}^0_{31}=0$
 with a representative set of the fixed
values of the parameters as
$\tan\beta =10, ~
 M_{\tilde{\xi}}= 120 ~\mbox{GeV}$,
 and found $\Omega_{\tilde{\xi}} h^2 \sim10^{-3}$,
where $\Omega_{\tilde{\xi}} h^2$  depends only very weakly
 on $m_0$ and $M_{1/2}$.
So,  for a wide range of mixing of
$\tilde{\eta}^{u0,d0}$ and $\tilde{\phi}$ the relic density $\Omega_{\tilde{\xi}} h^2$
is much smaller than the observed value.

If $M_{\tilde{\xi}}+M_{\tilde{\chi}} 
>  m_{\xi}$ is satisfied,
 where $m_{\xi}$
 is the mass of the lightest inert Higgs boson $\xi$ (either $\xi^0_{R1}$ or $\xi^0_{I1}$),
 then it is stable
despite (\ref{hierarchy}): Its decay is kinematically forbidden, and so
 it can be  a dark matter particle  (third dark matter), too
~\footnote{ Multi-component dark matter has been discussed e.g. in \cite{Cao:2007fy}.}.
 In Refs.~\cite{Barbieri:2006dq,LopezHonorez:2006gr} the feature of the inert Higgs boson dark matter was
studied in detail for non-SUSY models.  It turned  out \cite{LopezHonorez:2006gr} that to  obtain
a realistic relic density its mass  has to be either very small ($\lesssim 80$ GeV)
or very large  ($\gtrsim  500$ GeV) and the relic density 
is smaller than $  0.02$ between $100~\mbox{GeV}$ and $ 300~\mbox{GeV}$
\footnote{ The extra SUSY contributions, 
$ \xi+\xi\to \tilde{\chi}+\tilde{\chi},~ \tilde{\xi}+\tilde{\xi} $ etc., even decrease 
the relic density.}
(see also \cite{Araki:2011hm}).
As we will see, this range of $m_{\xi}$ is particularly interesting,
because we can expand   the allowed range in the $m_0$\,-\,$M_{1/2}$ plane considerably.

\begin{figure}[htbp]
\begin{center}
\includegraphics*[width=0.6 \textwidth]{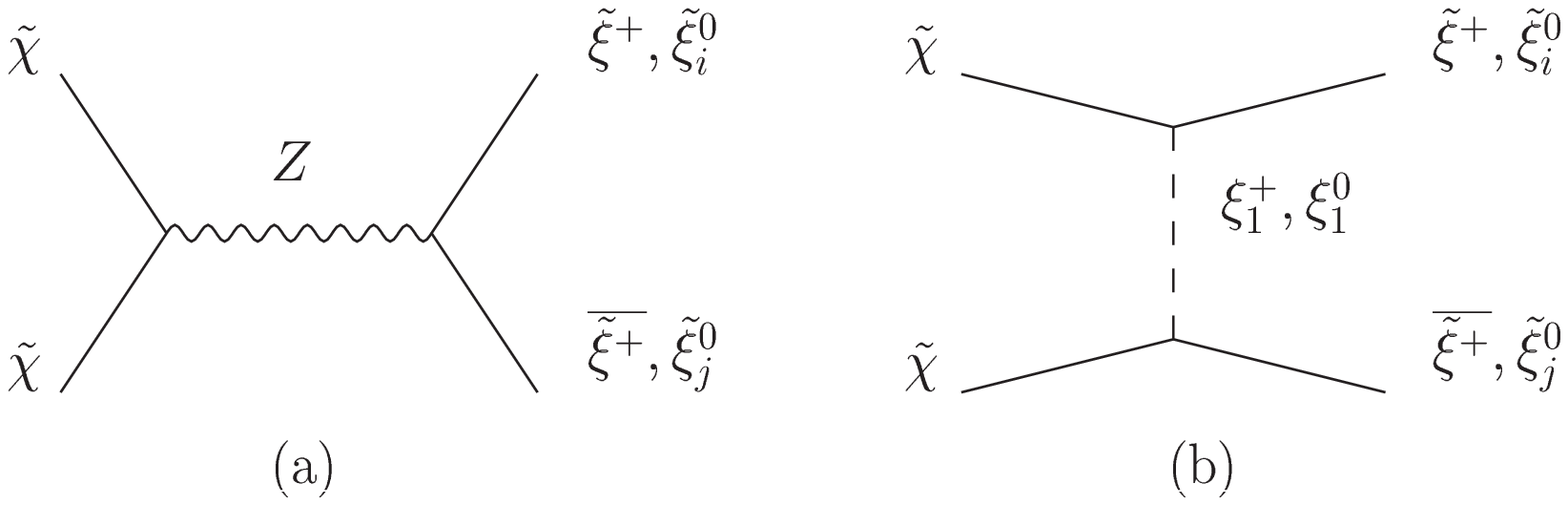}
\caption{\label{annihilation1}{\scriptsize Annihilation diagrams of
$\tilde{\chi}\tilde\chi\to \overline{\tilde{\xi}^+}\tilde{\xi}^+ ,~
\tilde{\xi}^0_i\tilde{\xi}^0_j $.}}
\end{center}
\end{figure}

Keeping this in mind we turn to the relic density of $\tilde{\chi}$.
As the assumption (\ref{hierarchy}) indicates 
$\tilde{\chi}$ dark matter particles can annihilate into  
$\tilde{\xi}^0_i \tilde{\xi}^0_j$ and $\overline{\tilde{\xi}^+} \tilde{\xi}^+$,
as shown in Fig.~\ref{annihilation1}, which contribute to the relic density $\Omega_{\tilde{\chi}} h^2$.
The MSSM part is the same as in the MSSM, so that
 for the case of the higgsino-like $\tilde{\chi}$ the dark matter mass $M_{\tilde{\chi}}$ 
should be ${\cal O}(1)$ TeV to obtain $\Omega_{\tilde{\chi}} h^2 =0.11$.
For a bino-like $\tilde{\chi}$, 
the contributions from the MSSM sector to
the annihilation cross section $\sigma_{\tilde{\chi}}$ is indeed too small.
But $\sigma_{\tilde{\chi}}$ can increase 
through the diagrams of Fig.~\ref{annihilation1} (b) in the inert Higgs sector.
This indicates that the parameter space of the MSSM sector  in this region can be
relaxed. This is what we would like to see below.
To this end we make further approximation:
\begin{figure}[tbp]
\begin{center}
\includegraphics*[width=0.9 \textwidth]{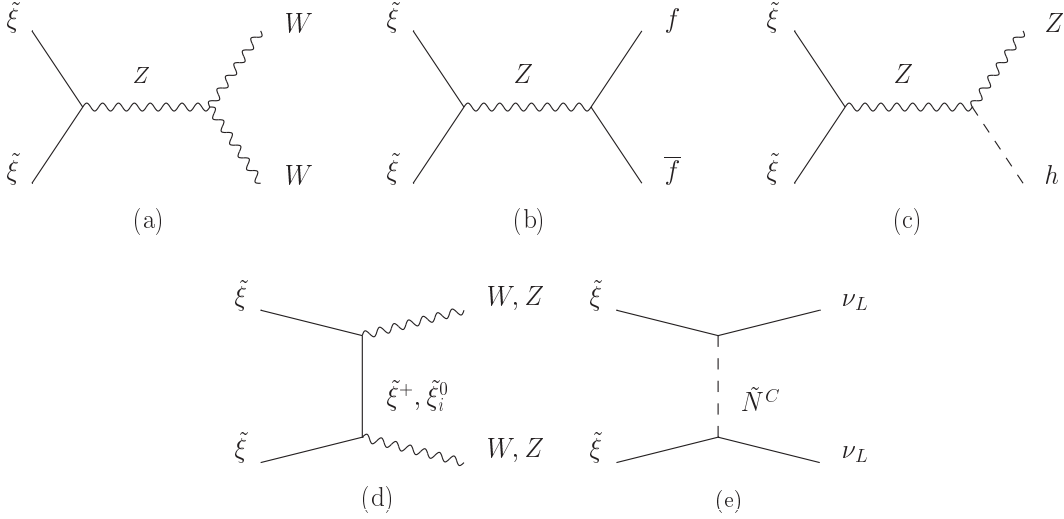}
\caption{\label{annihilation0}{\scriptsize Annihilation  diagrams of
$\tilde{\xi}$.}}
\end{center}
\end{figure}
As we see from Table III 
we  can always choose $B_\eta$
such that the contributions to the annihilation  mediated by
the exchange of the heavier ones are negligibly small
compared with those mediated by the exchange of the lighter ones.
Furthermore,
 $\xi^0_{R,I2} (\simeq \phi_{R,I})$ 
is not coupled with a gaugino-like $\tilde{\chi}$.
Therefore, the relevant part of the Lagrangian can be written as
\be
{\cal L}_{\rm eff} &=&\frac{g}{2}\left\{
\left(N_{12}+t_WN_{11}\right)
~\overline{\tilde{\chi}} ~\tilde{\xi}^+~(\xi_{1}^+)^*\nn
+\left(-N_{12}+t_W N_{11}\right)
\overline{\tilde{\chi}}[~(\tilde{C}^0_{1i})^*~ P_L+
\tilde{C}^0_{2i}~P_R~]\tilde{\xi}^0_i 
~\xi_1^0+h.c.\right\}\nn\\
& &-\frac{g}{2 c_W}\left\{ (1-2s_W^2)\overline{\tilde{\xi}^+}\gamma^\mu\tilde{\xi}^+
-\frac{1}{2} [ (\tilde{C}_{1i}^0)^* \tilde{C}_{1j}^0 - (\tilde{C}_{2i}^0)^* \tilde{C}_{2j}^0]~
\overline{\tilde{\xi}^0_i}\gamma^\mu P_L \tilde{\xi}^0_j\right.\nn\\
& &\left.~+\frac{1}{2} [ \tilde{C}_{1i}^0 (\tilde{C}_{1j}^0)^* -
 \tilde{C}_{2i}^0 (\tilde{C}_{2j}^0)^*]~
\overline{\tilde{\xi}_i^0}\gamma^\mu P_R \tilde{\xi}^0_j\right\}~Z_\mu~,
\label{Leff}
\ee
where $P_{{}^R_L}=(1\pm\gamma_5)/2$, 
$\xi_1^0=(\xi_{R1}^0+i \xi_{I1}^0)/\sqrt{2}, ~t_W=\tan\theta_W$,
$\tilde{C}^0_{ij}$ are defined in (\ref{c-tilde-xi0}), and 
the mixing parameters $N_{11} (N_{12})$ is
the bino (wino)  content in $\tilde{\chi}$, respectively.
To proceed we make use of the fact that
the relic density is inversely proportional to the annihilation
cross section, and approximate
the inverse of  $\Omega_{\tilde{\chi}} h^2$ as
\be
\frac{1}{\Omega_{\tilde{\chi}} h^2}
&\simeq&\frac{1}{\Omega_{\rm MSSM} h^2}+\frac{1}{\Omega_{\rm EXTRA} h^2}~,
\label{omega1}
\ee
where $\Omega_{\rm MSSM} h^2$ is the relic density of $\tilde{\chi}$ 
computed without including
the diagrams of Fig.~\ref{annihilation1}, while 
$\Omega_{\rm EXTRA} h^2$ is  computed only with  the annihilation processes 
of  Fig.~\ref{annihilation1}.
The approximation becomes better if one of 
$\Omega h^2$'s is dominated.
We compute  $\Omega_{\rm MSSM} h^2$  using
the package ``micrOMEGAs'' \cite{Belanger:2010gh}, and 
$\Omega_{\rm EXTRA} h^2$ analytically using the approximate formula
given in \cite{Griest:1988ma}.

According to \cite{Griest:1988ma} we expand the relativistic
cross section $\sigma (\tilde{\chi}\tilde{\chi}\to 
\overline{\tilde{\xi}^+}\tilde{\xi}^+ ,~
\tilde{\xi}^0_i\tilde{\xi}^0_j)$ in powers of their relative velocity $v$,
$ \sigma (\tilde{\chi}\tilde{\chi}\to 
\overline{\tilde{\xi}^+}\tilde{\xi}^+ ,~
\tilde{\xi}^0_i\tilde{\xi}^0_j)v =a_{\rm EXTRA}+b_{\rm EXTRA} v^2 $. 
We then use \cite{Griest:1988ma} 
\be
\Omega_{\rm EXTRA} h^2 &\simeq &  2.82 \times 10^8
 \left[ \frac{M_{\tilde{\chi}}}{\mbox{GeV}} \right] Y_\infty~,
\ee
where
\be
Y_\infty^{-1} &=&
0.264 M_{\rm PL} 
M_{\tilde{\chi}} g_*^{1/2}(a_{\rm EXTRA}/x_f+3 b_{\rm EXTRA}/x_f^2)~,
\label{omaga2}
\ee
$M_{\rm PL}=
1.22\times 10^{19}$ GeV 
is the Planck mass, $g_*
\simeq 90$ and $x_f=M_{\tilde{\chi}}/T$ is 
the inverse dimensionless freeze-out temperature. 
The quantities $a_{\rm EXTRA}$ and $b_{\rm EXTRA}$ are found to be
\be
a_{\rm EXTRA} &=&  
\sum_{\mbox{\tiny{$\overline{\tilde{\xi}^+}\tilde{\xi}^+ ,~ 
\tilde{\xi}^0_i\tilde{\xi}^0_j $}}}
\frac{4G_F^2 m_W^4 
\sqrt{1-M_{\tilde{\xi}}^2/M_{\tilde{\chi}}^2}}{\pi\left(m_\xi^2-
M_{\tilde{\xi}} ^2+M_{\tilde{\chi}}^2\right)^2} 
\left( M_{\tilde{\xi}}^2(w_1+w_4)+w_2 M_{\tilde{\chi}}^2+w_3 M_{\tilde{\xi}}
 M_{\tilde{\chi}}\right)~,
\label{a-extra}\\
b_{\rm EXTRA} &=&
\sum_{\mbox{\tiny{$\overline{\tilde{\xi}^+}\tilde{\xi}^+ ,~ 
\tilde{\xi}^0_i\tilde{\xi}^0_j$}}}
\left\{
\frac{G_F^2 m_W^4\sqrt{1-M_{\tilde{\xi}} ^2/M_{\tilde{\chi}}^2}}
   {6 \pi\left(M_{\tilde{\chi}}^2-M_{\tilde{\xi}}^2\right) 
   \left(m_\xi^2-M_{\tilde{\xi}}^2+M_{\tilde{\chi}}^2\right) ^4}
\sum_{i=1}^{4} w_i B_i
\right.\nn\\
&&\left. +
\frac{2 G_F^2 m_Z^4\sqrt{1-M_{\tilde{\xi}} ^2/M_{\tilde{\chi}}^2}
(N_{13}^2-N_{14}^2)^2(M_{\tilde{\xi}}^2+ 2 M_{\tilde{\chi}}^2)}
   {3 \pi\left( 16 M_{\tilde{\chi}}^4-8M_{\tilde{\chi}}^2 m_Z^2+m_Z^4+\Gamma_Z^2 m_Z^2\right)}~w_5
\right\}~,
\label{b-extra}\\
B_1 &=&
(M_{\tilde{\xi}}^2-M_{\tilde{\chi}}^2)^2(13M_{\tilde{\xi}}^4-
10 M_{\tilde{\chi}}^2 M_{\tilde{\xi}}^2+16 M_{\tilde{\chi}}^4)
+m_\xi^4(13M_{\tilde{\xi}}^4-26 M_{\tilde{\chi}}^2 M_{\tilde{\xi}}^2+16 M_{\tilde{\chi}}^4)\nn\\
& &+m_\xi^2(-26M_{\tilde{\xi}}^6+
70 M_{\tilde{\chi}}^2 M_{\tilde{\xi}}^4-44M_{\tilde{\chi}}^4 M_{\tilde{\xi}}^2)~,
\label{b1-extra}\\
B_2 &=&
(M_{\tilde{\xi}}^2-M_{\tilde{\chi}}^2)^2(6M_{\tilde{\xi}}^4+
17 M_{\tilde{\chi}}^2 M_{\tilde{\xi}}^2-4 M_{\tilde{\chi}}^4)
+3m_\xi^4(2M_{\tilde{\xi}}^4-5 M_{\tilde{\chi}}^2 M_{\tilde{\xi}}^2+4 M_{\tilde{\chi}}^4)\nn\\
& &-2m_\xi^2(6M_{\tilde{\xi}}^6-13 M_{\tilde{\chi}}^4 M_{\tilde{\xi}}^2
-M_{\tilde{\chi}}^2 M_{\tilde{\xi}}^4
+8 M_{\tilde{\chi}}^6)~,
\label{b2-extra}\\
B_3 &=&
M_{\tilde{\xi}} M_{\tilde{\chi}}\left[ 3 m_\xi^4 M_{\tilde{\xi}}^2
+19 M_{\tilde{\xi}}^2(M_{\tilde{\xi}}^2-M_{\tilde{\chi}}^2)^2\!\!+\!
m_\xi^2(-22M_{\tilde{\xi}}^4+62 M_{\tilde{\chi}}^2 M_{\tilde{\xi}}^2-40 M_{\tilde{\chi}}^4)\right],
\label{b3-extra}\\
B_4 &=&
M_{\tilde{\xi}}^2\left[(21 M_{\tilde{\xi}}^2-2 M_{\tilde{\chi}}^2)(M_{\tilde{\xi}}^2-M_{\tilde{\chi}}^2)^2
-3 m^4_\xi(6 M_{\tilde{\chi}}^2-7 M_{\tilde{\xi}}^2)\right.\nn\\
& &\left.-6m^2_\xi(7 M_{\tilde{\xi}}^4-17 M_{\tilde{\chi}}^2 M_{\tilde{\xi}}^2+ 10 M_{\tilde{\chi}}^4 )\right]~,
\label{b4-extra}
\ee
where for the annihilation processes $\tilde\chi\tilde\chi\to\overline{\tilde{\xi}^+} \tilde{\xi}^+$
\be
w_1&=&
\frac{1}{32}\left(N_{12}+t_WN_{11}\right)^4=\frac{w_2}{2}=\frac{w_3}{4}=w_4~,~
w_5 =\frac{1}{16}(1-2 s_W^2)^4~,
\ee
and for $\tilde\chi\tilde\chi\to\tilde{\xi}^0_i \tilde{\xi}^0_j$
\be
w_1&=&
\frac{1}{32}\left(-N_{12}+t_WN_{11}\right)^4
=\frac{w_2}{2}=\frac{w_3}{4}=w_4~,~
w_5 =\frac{1}{16}.
\ee
To simplify the situation we have assumed
  that all the inert higgsinos have the same mass as $\tilde{\xi}$  
(i.e. $M_{\tilde{\xi}}= M_{\tilde{\xi}^0_1}\simeq  M_{\tilde{\xi}^0_2}\simeq  
M_{\tilde{\xi}^0_3}\simeq  
M_{\tilde{\xi}^+}$)  and 
the light inert Higgs bosons also have the common mass 
  $m_\xi$ (i.e. $m_\xi =m_{\xi^0_{R1}} \simeq m_{\xi^0_{I1}} 
  \simeq m_{\xi^+_1}$).
In this limit the mixing parameters can be approximated
as   $\tilde{C}^0_{1i}=(i,1,0)/\sqrt{2}~,~
\tilde{C}^0_{2i}=(-i,1,0)/\sqrt{2}$.

%%%
  \begin{figure}[htbp]
\begin{center}
\includegraphics*[width=13cm]{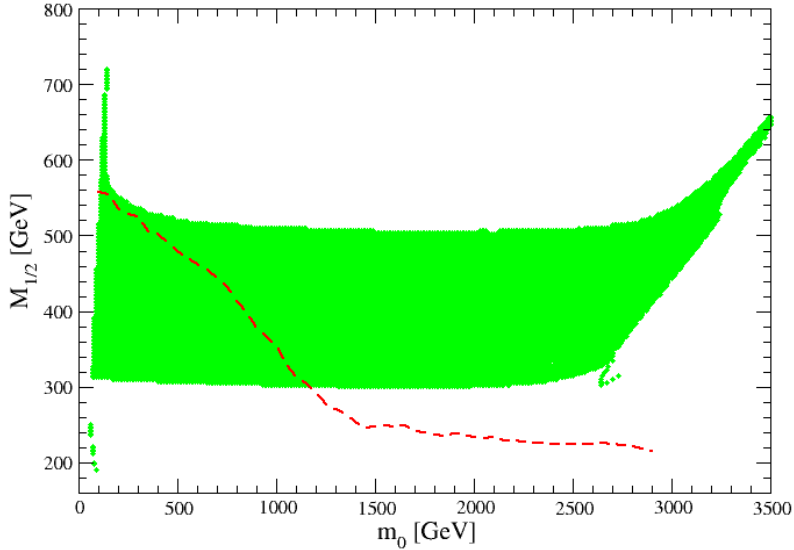}
\caption{\label{omega-chi}{\scriptsize
The allowed region in the $m_0$\,-\,$M_{1/2}$ plane
    for  $m_\xi > M_{\tilde{\chi}} > M_{\tilde{\xi}} = 120~\mbox{GeV}$ with
    $A_0=0,~\tan\beta=10$ and $\mbox{sign}(\mu_H)=+$, where we have used:
$\sin^2\theta_W=0.23,~m_W=80.4~\mbox{GeV},~  G_F=1.17\times
10^{-5}~\mbox{GeV}^{-2}$.
    In the actual calculation of  $\Omega_{\tilde{\chi}} h^2$, we have approximated
  that all the inert higgsinos have the same mass as $\tilde{\xi}$  
(i.e. $M_{\tilde{\xi}}= M_{\tilde{\xi}^0_1}\simeq  M_{\tilde{\xi}^0_2}\simeq  
M_{\tilde{\xi}^0_3}\simeq  
M_{\tilde{\xi}^+}$)  and the light inert Higgs bosons also have the common mass 
  $m_\xi$ (i.e. $m_\xi= m_{\xi^0_{R1}} \simeq m_{\xi^0_{I1}} 
  \simeq m_{\xi^+_1}$).
 The constraints coming from the stau LSP, the electro-weak symmetry breaking
and the  LEP chargino mass limit are included,  and
the dashed line is the recent LHC limit \cite{Aad:2011ib}.  (See the comment of footnote 3.)}.}
\end{center}
\end{figure}

\begin{figure}[tbp]
\begin{center}
\includegraphics*[width=10cm]{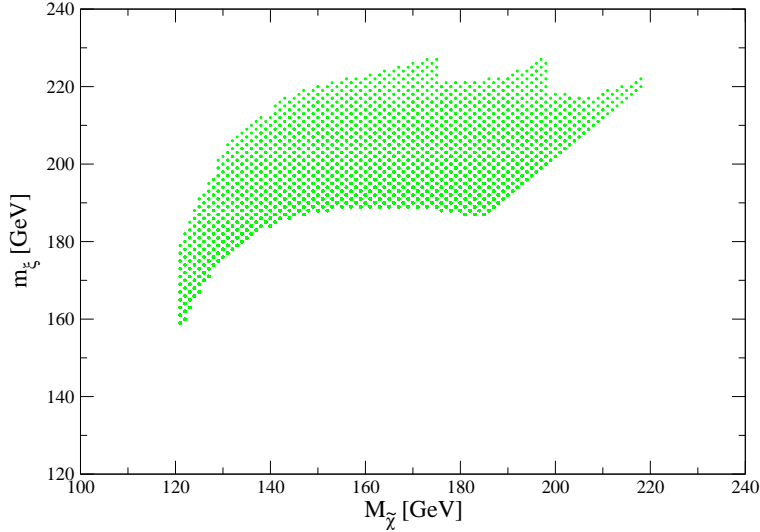}
\caption{\label{mchi-mxi}{\scriptsize
The region
in the $M_{\tilde{\chi}}$\,-\,$m_\xi$ plane for $M_{\tilde{\xi}}=120$ GeV,
which gives the green  area of Fig.~\ref{omega-chi} 
except the area along the left and right border lines.}
}
\end{center}
\end{figure}

 Now we present  the calculation of $\Omega_{\tilde{\chi}} h^2$.
 In Fig.~\ref{omega-chi} we plot the allowed region in the 
  $m_0$\,-\,$M_{1/2}$ plane. 
 We have obtained the allowed region
  for  $m_\xi > M_{\tilde{\chi}} > M_{\tilde{\xi}} = 120~\mbox{GeV}$ with $\tan\beta =10$.
The allowed region in Fig.~\ref{omega-chi} should be compared with 
that in Fig.~\ref{cmssm}, which is obtained under the assumption that
the annihilation cross section of $\tilde{\chi}$ into the 
inert Higgs sector is sufficiently suppressed.
 The mass values of  $m_\xi$ and $ M_{\tilde{\chi}}$ giving  
the green area of Fig.~\ref{omega-chi}, except the area along the 
left and right border lines, 
 are shown in Fig.~\ref{mchi-mxi}. 
($M_{\tilde{\xi}}$ is fixed at $120~\mbox{GeV}$.)
We see that $m_\xi $ is smaller than $M_{\tilde{\chi}} + M_{\tilde{\xi}}$  
in this area  so that 
 the lightest neutral inert Higgs boson is a dark matter particle, too.
As we mentioned, the feature of the inert Higgs boson dark matter was
studied in Refs.~\cite{Barbieri:2006dq,LopezHonorez:2006gr}.
In our case $m_\xi$ varies from $160 $ to $ 220$ GeV (see Fig.~\ref{mchi-mxi}).
Using their results, we find that the contribution 
of the inert Higgs boson dark matter to the relic density $0.11$ is at most  $15\%$,
where the extra SUSY contributions, 
$ \xi+\xi\to \tilde{\chi}+\tilde{\chi},~ \tilde{\xi}+\tilde{\xi} $ etc., are not
taken into account.
These extra contributions can be as large as that without them.
We roughly estimate that dark matter
consists more than $90 \%$ of  $\tilde{\chi}$ in the most of the green area
of Fig.~\ref{omega-chi}.
The area along the  left and right border lines is the area very closed to
the red line of Fig.~\ref{cmssm}.
 In this area 
$m_\xi > M_{\tilde{\chi}} + M_{\tilde{\xi}}$  
is satisfied, and 
 the annihilation cross section for 
$\tilde{\chi}\tilde{\chi}\to 
\overline{\tilde{\xi}^+}\tilde{\xi}^+ ,~\tilde{\xi}^0_i\tilde{\xi}^0_j$
 is very small, implying that dark matter consists almost 100 percent of  $\tilde{\chi}$
 in this area.

If the masses $m_\xi $, $ M_{\tilde{\chi}}$ and $M_{\tilde{\xi}}$ are very close, 
there can be co-annihilations among the dark matter particles.
We see from Fig.~\ref{mchi-mxi}  that
there exists a small region where $ M_{\tilde{\chi}} \simeq  M_{\tilde{\xi}}$ 
and  $ m_{\xi} \simeq  M_{\tilde{\chi}}$  are satisfied, respectively.
For these regions, co-annihilation processes such as
 $\tilde{\chi}+\tilde{\xi}\to \xi_{R1}^0+Z, ~
 \tilde{\chi}+\xi_{R1}^0 \to \tilde{\xi}+Z$ may become possible.
 Indeed, $ m_{\xi}$ and $ M_{\tilde{\chi}}$ are degenerate in
the area close to the upper borderline of the green  region in Fig.~\ref{omega-chi},
 while $ M_{\tilde{\xi}}$ and $ M_{\tilde{\chi}}$ are degenerate in
 the area close to the lower borderline.
So, one should take into account the effects of the co-annihilation
processes. However,  we  have ignored them in Fig.~\ref{omega-chi},
because these effects will change only the narrow area 
close to the borderlines and not the gross structure of the allowed region
in  Fig.~\ref{omega-chi}.

 There will be  some differences in direct and indirect searches of dark matter.
Let us make a few comments on this, where the details will be published elsewhere.
We recall that the direct rate is proportional to the relic density, while the indirect 
rate is  proportional to the square of the relic density. 
Therefore,  indirect search of the inert Higgs boson dark matter
suffers from  a suppression factor of at least $0.1^2$
($\sim 10^{-6}$ in the case of the inert higgsino dark matter).
At first sight, indirect detection rate of the dark matter $\tilde{\chi}$ 
seems to be   suppressed compared with the case of the CMSSM, because
the higgsino portion of $\tilde{\chi}$ is very small 
in the green area of Fig.~\ref{omega-chi} except on the left and right border lines.
However,  annihilation  not only into the  neutral $\tilde{\xi}$'s, 
but also into the charged $\tilde{\xi}^+$'s is possible,  and this rate is large.
So, indirect detection  of $\tilde{\chi}$ has to be carefully studied.

Direct detection of  $\xi$ 
suffers from  a suppression factor of at least $0.1$.
Using the result of \cite{LopezHonorez:2006gr} we may conclude that
direct detection of  $\xi$ in the mass range in question, i.e.
 $160 ~\mbox{GeV}\lsim m_{\xi} \lsim 220~\mbox{GeV}$,
does not need to be discussed.
As for direct detection of  $\tilde{\chi}$ 
the spin-dependent cross section with the nuclei  is much smaller 
than in the case of the CMSSM, because
the higgsino portion of $\tilde{\chi}$ is small 
in the most of area of the green region of Fig.~\ref{omega-chi}.
Instead,  the spin-independent cross section with the nuclei 
 is of the same order as in the case of the MSSM.
The relatively large coupling of $\tilde{\chi}$ to the inert Higgs sector does not 
change  the cross section with the nuclei.

So our conclusion is 
that  the allowed region in the $m_0$\,-\,$M_{1/2}$ plane is considerably enlarged, if the inert higgsinos are
lighter than $\tilde{\chi}$,
 although the dominant component {\bf ($\gsim 90\%$)} of dark matter is $\tilde{\chi}$.
 There is a wide allowed region even above the LHC limit 
 (dashed line in Fig.~\ref{omega-chi})  \cite{Aad:2011ib}.
Direct and indirect searches of dark matter are
slightly different compared with the cases of the CMSSM.
We will leave the analysis for our future project.

At last we would like to emphasize that the radiative seesaw model of Ma 
\cite{Ma:2006km} is a two-Higgs-doublet model with a
specific structure of the Higgs sector (see e.g. 
\cite{Barbieri:2006dq,LopezHonorez:2006gr,Cao:2007rm}
and also \cite{Branco:2011iw}), which 
is intimately related to the neutrino mass and mixing and
indirectly to the lepton flavor violations.
Furthermore, in supersymmetric case, the Yukawa terms
$\lambda^{u(d)} {\bf H}^{d(u)} \mbox{\boldmath  $ \eta $}^{u(d)} 
\mbox{\boldmath  $ \phi $}$  along with the corresponding
A-terms will contribute to the radiative correction to the Higgs mass
\cite{Okada:1990vk} so that the  upper bound on the  lightest Higgs mass will change.
In the light of the recent  LHC results \cite{higgs} 
the computation of the upper bound is
important for the model, although it is not directly related to the
problem of dark matter.
Therefore, more detailed investigations of the Higgs sector
of the present model
along this line will be included to our future study.

\vspace*{5mm}
M.~A. and J.~K. are partially supported by a Grant-in-Aid for Scientific
Research (C) from Japan Society for Promotion of Science (Nos. 22740137
and 22540271, respectively).
M.~A. is  partially supported by the Hokuriku bank.

\end{document}